# LATERAL IBIC ANALYSIS OF A 4H-SiC SCHOTTKY DIODE

P. Olivero[1,2], J. Forneris[1], P. Gamarra[1], M. Jakšić[2], A. Lo Giudice[1], C. Manfredotti[1],

Ž. Pastuović[2], N. Skukan[2], E.Vittone[1#]

[1] Experimental Physics Dept./ NIS Excellence Centre, University of Torino, and INFN-Sez. di Torino via P.Giuria 1, 10125 Torino, Italy

[2] Ruđer Bošković Institute, Bijenicka 54, P.O. Box 180, 10002 Zagreb, Croatia

**Abstract**

The transport properties of a 4H-SiC Schottky diode have been investigated by the Ion Beam Induced Charge Collection (IBIC) technique in lateral geometry through the analysis of the charge collection efficiency (CCE) profile at a fixed applied reverse bias voltage. The cross section of the sample orthogonal to the electrodes was irradiated by a rarefied 4 MeV proton microbeam and the charge pulses have been recorded as function of incident proton position with a spatial resolution of 2 µm. The CCE profile shows a broad plateau with CCE values close to 100% occurring at the depletion layer, whereas in the neutral region, the exponentially decreasing profile indicates the dominant role played by the diffusion transport mechanism. Mapping of charge pulses was accomplished by a novel computational approach that consists in mapping the Gunn's weighting potential by solving the electrostatic problem by finite element method and hence evaluating the induced charge at the sensing electrode by a Monte Carlo method. The combination of these two computational methods enabled an exhaustive interpretation of the experimental profiles and allowed an accurate evaluation of both the electrical characteristics of the active region (e.g. electric field profiles) and of basic transport parameters (i. e. diffusion length and minority carrier lifetime).

---

[#] Correspondig author; Ettore Vittone, Dipartimento di Fisica Sperimentale, Università di Torino, via Pietro Giuria 1, 10125 Torino (Italy), Tel. +39-0116707371, Telefax +39-0116691104; e-mail:ettore.vittone@unito.it.

## 1. Introduction

In the last 17 years, the measurement of the charge induced at sensitive electrodes in consequence of the irradiation of samples with light (mainly H or He) ions was successfully applied for the evaluation of the transport properties and the electrostatic features of numerous semiconductor basic devices, such as junction or Schottky diodes, solar cells, MOS diodes and to probe the effects of heavy ion irradiation on solid state detectors [1]. In order to exploit the full analytical capabilities of this technique, a theory based on fundamental electrostatic theorems and lemmas has been developed and validated by benchmark experiments [2, 3]. All these experiments have been carried out in frontal geometry (i.e. irradiating the sensitive electrode) and the results have been interpreted by comparing the experimental profiles and images with maps of charge pulses accomplished by solving numerically the adjoint carrier continuity equations [4] by the finite element method [3].

The purpose of this paper is to present another approach to the problem of the evaluation of the induced charge, based on a simple, one-dimensional Monte Carlo algorithm. The experiment consists in the lateral IBIC analysis carried out for the first time on the epitaxial layer of a 4H-SiC Schottky diode. This experimental approach allows a direct and meaningful insight on the mechanisms underlying the formation of the induced charge pulses, with a clear discrimination of the different contributions deriving from the charge generation in the depletion and neutral regions. The Monte Carlo method presented in this paper simulates the transport of carriers in the active region and takes into account both diffusion and drift mechanisms.

## 2. Experimental

The sample under test is a Schottky diode fabricated by Alenia Marconi on an n-type (~50 μm thick) epitaxial layer produced by the Institut für Kristallzüchtung (IKZ), Berlin on a $n^+$ substrate with low micropipe density (16-30 $cm^{-3}$) produced by CREE. The Schottky and ohmic electrodes (1.5 mm diameter) were made by Alenia Marconi Systems by evaporating Ni/Au on the epitaxial layer and a Ti/Pt/Au on the substrate side (C-face), respectively [5].

The diode was cleaved in order to expose the cross section to the ion beam irradiation. The cleavage did not remarkably increase the leakage current, which was below 1 nA at a reverse bias voltage of 100 V.

Electrical characterization was carried out by means of current/voltage and capacitance/voltage measurements providing an ideality factor of 1.1, a zero voltage barrier height of 1.5 eV and an average doping profile of $1.5 \cdot 10^{14}$ donors $cm^{-2}$. The sample was then edge-on mounted in order to perform lateral IBIC measurements (Fig. 1), which were carried out at the microbeam facility of the Ruđer Bošković Institute of Zagreb (HR) using 4 MeV protons. The beam was focused and scanned over the cleaved cross section of the SiC diode with an ion current of $<10^3$ protons $s^{-1}$ in order to avoid electronic pile-up and to reduce the radiation damage. From the analysis of proton transmission profiles of a copper grid (repeat size of 25 μm), a beam spot size of 2 μm was evaluated by the FWHM of the profiles assumed to be gaussian.

The proton beam was scanned on the cross section of the diode (100×100 $\mu m^2$ scan area) from the Schottky contact to the highly doped substrate. The range of 4 MeV protons in SiC is about 100 μm, as evaluated by the SRIM code. Since the electron/hole generation occurs primarily at the Bragg's peak, in the following analysis we will neglect any recombination at the irradiated surface. Moreover, because of the low fluence, we will also neglect the built-in electric field due to the trapping of carriers at the incident proton surface. Under these hypotheses, the results we present here can be interpreted in a one-dimensional model, i.e. considering the motion of carriers along the y direction which is parallel to the electric field and normal to the electrode surface.

The electronic chain comprises a CANBERRA 2004 charge sensitive preamplifier, a CANBERRA 2021 amplifier (shaping time=1.5 µs). A home-made data acquisition hardware and software was used to acquire and store every IBIC pulse along with its spatial coordinates [6].

The calibration of the electronic chain was performed by using a Si surface barrier detector and a precision pulse generator, in order to relate pulse heights provided by the reference Si detector (for which 100% charge collection efficiency was assumed) with those from the SiC diode. Assuming an average energy to create electron/hole pairs of 3.62 eV and 7.78 eV in Si and SiC, respectively [5], the spectral sensitivity of the IBIC set-up was (1800±10) electrons, corresponding to about 0.3% of CCE for 4 MeV protons in 4H-SiC.

Fig. 2 shows the lateral IBIC map obtained at a reverse bias voltage of 50 V. The CCE is encoded in the colour scale, which represents the median of the IBIC pulse distribution for each pixel. Fig. 3 shows the CCE profile obtained by projecting the IBIC map on the y axis. The continuous line indicates the median of the pulse distribution evaluated at each y position. On the right side, the CCE spectra evaluated at four different positions are shown as an example.

The behaviour of the CCE profile resembles what was reported in a previous paper for a Si $p^+n$ junction diode [7]. The plateau at about 100% CCE extends from 0 to around 20 µm and corresponds to the depletion region, defined as the region where the Gunn's weighting potential is not null [8] and a strong electric field occurs. The fast drift allows carriers to cross the active region in times much smaller than their lifetime, hence, the charge induced at the electrode is entirely collected.

The monotonically decreasing behaviour of the CCE profile at deeper (i.e. for y>20 µm) locations can be interpreted as due to the contribution of minority carriers (holes) generated by protons in the neutral region, and injected by diffusion into the depletion region, where the IBIC signal is formed. The exponential-like behaviour accounts for the probability for holes to reach the edge ($w_d$) of the depletion layer; in fact, if constant lifetime and diffusivity are assumed, the carrier concentration varies according to expression $\exp[-(x-w_d)/L_p)$, where $L_p$ is the hole diffusion length [8]. The exponential fit of the right side of the median curve, provides a diffusion length of (4.4±0.7) µm.

## 3. Theoretical interpretation

A more rigorous interpretation of the experiment was carried out in the framework of the theory based on the Shockley-Ramo-Gunn theorem, described in several previous papers [1, 8]. In summary, a charge moving in a region with an arbitrary arrangement of charges and conductors held at fixed potential induces a current flowing from ground to one electrode connected to the electronic amplification chain (i.e. the sensitive electrode).

If an infinitesimal charge q is located at position $r_0$ at the starting time and, after a certain time T, is at position r, then the induced charge at the sensitive electrode held at constant potential V can be evaluated as the difference of the Gunn's weighting potential between the initial and final position, i.e. [1, 9]:

$$1) \quad Q = q\left[\left.\frac{\partial \psi}{\partial V}\right|_r - \left.\frac{\partial \psi}{\partial V}\right|_{r_0}\right]$$

Where $\psi$ is the electrostatic potential evaluated by solving the relevant Poisson's equation, and considering the suitable boundary conditions (i.e. the sensitive electrode at potential V and all the other conductors at fixed potential). As a consequence, to evaluate the charge induced by the motion of free carriers generated in a certain point $r_0$, we have adopted the following three-step model:

i) solution of the electrostatic problem (i.e. the calculation of the electrostatic potential profile) assuming a one-dimensional geometry; the electrostatic domain is defined as the 4H-SiC epitaxial layer, with boundary conditions defined by the contact potential summed to the applied bias voltage at the Schottky electrode and the ground potential at the highly doped substrate.

ii) Calculation of the weighting potential approximating $\frac{\partial \psi}{\partial V}$ as a central finite difference, i.e. $\frac{\partial \psi}{\partial V} \cong \frac{\psi(r, V+\Delta V) - \psi(r, V-\Delta V)}{2 \cdot \Delta V}$ (in our calculations, $\Delta V$ was arbitrarily set to 0.001 V).

iii) Calculation of the position r of a point charge, generated at time t=0 in $r_0$, at time T and evaluation of the induced charge through eq. 1).

The electrostatic problem (i.e. items i) and ii)) was modeled using finite-element code COMSOL Multiphysics 3.5a® software package [10], as described in [3]. The Gunn's weighting potential is shown in Fig. 3.

The problem of the evaluation of the motion of the carriers has been faced with a novel method based on the Monte Carlo solution of the transport equations of free carriers [11].

Considering holes, the continuity equation is:

2) $\frac{\partial p}{\partial t} = \frac{\partial}{\partial y}\left[D_p \cdot \frac{\partial p}{\partial y} - \mu_p \cdot E \cdot p\right] - \frac{p}{\tau_p}$

Where p is the hole concentration, E is the electric field and $D_p$, $\mu_p$, $\tau_p$ are the diffusivity, mobility and lifetime, respectively. Such equation can be solved using the probabilistic interpretation of its explicit finite difference form:

$$p(y, t+\Delta t) = p(y,t) \cdot \left[ 1 - \frac{2\Delta t}{\Delta y^2} \cdot D_p(y) - \Delta t \cdot \frac{d(\mu_p \cdot E)}{dy} - \frac{\Delta t}{\tau_p} \right] +$$

$$3) \quad + p(y+\Delta y, t) \cdot \left[ \frac{\Delta t}{\Delta y^2} \cdot D_p(y) + \frac{\Delta t}{2 \cdot \Delta y} \cdot \left( \frac{dD_p}{dy} - \mu_p \cdot E \right) \right] +$$

$$+ p(y-\Delta y, t) \cdot \left[ \frac{\Delta t}{\Delta y^2} \cdot D_p(y) - \frac{\Delta t}{2 \cdot \Delta y} \cdot \left( \frac{dD_p}{dy} - \mu_p \cdot E \right) \right]$$

Which was obtained by superimposing a one-dimensional grid of constant mesh size $\Delta y$ and replacing the spatial first and second derivatives by central difference in y, and the derivative with respect to time by a first order forward difference. After defining $y_j \equiv j \cdot \Delta y$ and $t_n \equiv n \cdot \Delta t$, eq. 3) can be re-arranged as follows:

$$p(j, n+1) = p(j,n) \cdot P_0(j,n) + p(j+1, n) \cdot P_-(j,n) + p(j-1,n) \cdot P_+(j,n),$$
where

$$4) \quad P_0(j,n) = \left[ 1 - \frac{2\Delta t}{\Delta y^2} \cdot D_p(n) - \Delta t \cdot \frac{d(\mu_p \cdot E)}{dy} \bigg|_{y=n \cdot \Delta y} - \frac{\Delta t}{\tau_p} \right]$$

$$P_-(j,n) = \left[ \frac{\Delta t}{\Delta y^2} \cdot D_p(n) + \frac{\Delta t}{2 \cdot \Delta y} \cdot \left( \frac{dD_p}{dy} \bigg|_{y=n \cdot \Delta y} - (\mu_p \cdot E)_{y=n \cdot \Delta y} \right) \right]$$

$$P_+(j,n) = \left[ \frac{\Delta t}{\Delta y^2} \cdot D_p(n) - \frac{\Delta t}{2 \cdot \Delta y} \cdot \left( \frac{dD_p}{dy} \bigg|_{y=n \cdot \Delta y} - (\mu_p \cdot E)_{y=n \cdot \Delta y} \right) \right]$$

The probabilistic interpretation of eq. 4) is the following: let $p(j,n)$ be the probability that after n time steps, one carrier can be found at position $y=j \cdot \Delta y$; similarly $p(j+1,n)$ and $p(j-1,n)$ are the probabilities that a carrier is at point $y=(j+1) \cdot \Delta y$ and $y=(j-1) \cdot \Delta y$, respectively.

Hence:

- $P_0(j,n)$ is the probability that the particle remains at the same position during the time interval $\Delta t$;

- $P_+$ is the probability of the particle moving a step to the right from $y=(j-1) \cdot \Delta y$ to $y=j \cdot \Delta y$;

- $P_-$ is the probability of the particle moving a step to the left from $y=(j+1) \cdot \Delta y$ to $y=j \cdot \Delta y$.

Finally, the probabilistic interpretation of the coefficients $P_{0,+,-}$ is valid only if their sum is less or equal to 1 and if they are positive; this limitation determines the maximum dimensions of the space-time step [12]. Similar expressions can be easily found for electrons.

The simulation is then carried out by following the random walk of a predefined population of carriers generated at different points distributed through the epitaxial layer, evaluating their position after N time steps and hence calculating the induced charge through eq. 1).

The simulation of the CCE profile is then carried out for each nominal beam position, by following these steps:

i) Once the nominal beam position is chosen, $N_P$ generation points are extracted from a Gaussian distribution with a dispersion (FWHM=4.4 μm) evaluated by the convolution of the lateral straggling of protons (3.9 μm) combined in quadrature with the beam spot size (2 μm).

ii) At any generation point, the random walk of $N_C$ electrons and holes is followed and their position is evaluated after a predefined integration time T. For each carrier, the induced charge is calculated through eq. 1). The average of the $N_C$ induced-charge values is then representative of the mean CCE from a single incident ion.

iii) The $N_P$ simulated ions provide a distribution of CCE's which is relevant for the chosen nominal beam position, from which the median or other statistical parameters can be extracted.

The physical parameters used for the simulation were a mobility of 115 cm$^2$ s$^{-1}$ V$^{-1}$ and 800 cm$^2$ s$^{-1}$ V$^{-1}$ for electrons and holes respectively [13,3], and a fixed lifetime of 80 ns.

We have simulated the CCE profile using $N_P$=400 ions distributed through the thickness of the diode with a constant step of 0.44 μm. The CCE for each ion has been evaluated by following the random walk of $N_C$=100 carriers and considering an integration time sufficient to assure the saturation of the signal (i.e. of the order of 200 ns). Moreover, the electronic noise has been taken into account by adding, for each ion, a Gaussian fluctuation with a FWHM of 1% and a threshold of 7%.

Fig. 4 shows the excellent agreement of the median CCE evaluated with the above-described Monte Carlo simulation with the corresponding experimental profile. Moreover, the simulated pulse dispersions reported on the right side of Fig. 4 are qualitatively very similar to the the corresponding experimental ones, which are reported on the right side of Fig. 3 for beam positions of 5, 19, 21 and 25 µm from the Schottky electrodes. In order to quantitatively estimate the accuracy of the Monte Carlo simulation with respect to the experimental data not only in terms of the median profile but also of the pulse distribution, we have evaluated, for each nominal beam position, the difference between the $3^{rd}$ and $1^{st}$ quartile for both the experimental and simulated pulse distributions. Also in this case, the agreement between numerical and exoerimental data is excellent, as shown in Fig. 5.

The behaviour of this curve is strictly related to the slope of the collection efficiency curve. In the depletion region the CCE spectra are very narrow, and the flat spread profile is determined by the electronic noise. In correspondence of the edge of the depletion region (i.e. positions 2 and 3 of Fig. 3), the pulse distribution shows a tail at lower CCE values due to the free carrier generation within the neutral region. A broader distribution occurs when the entire carrier generation occurs into the neutral region (i.e. posizion 4 of Fig. 3) and its shape results from the convolution of the generation profile along the y direction and the exponential decay of the charge collection profile.

## 4. Conclusions

In this paper we present the analysis of a lateral IBIC experiment carried out on a 4H-SiC Schottky diode. The novel Monte Carlo approach used to simulate the pulse shape formation is based on the probabilistic interpretation of the explicit finite difference approximation of the continuity equations, which are solved by the random walk method. For each of the $N_P$ nominal generation positions, whose distruibution is modelled as Gaussian to take into account proton straggling and beam size, the mean induced charge was evaluated by simulating the transport of $N_C$ electrons and holes, each of them providing a value of the induced charge through eq. 1). The median of the distribution of the simulated $N_P$ pulses was evaluated in correspondence of each nominal beam position and the relevant CCE profile is in excellent agreement with the experimental data. Similarly good agreement has been obtained by the comparison of the simulated and experimental CCE spread profiles.

The Monte Carlo algorithm was proven to be robust, easy to be implemented in a computer programme and suitable to be developed for parallel computing. Although this preliminary work was limited to the analysis of a simple one-dimensional case of study, no difficulties arise for the extension of the model to higher dimensions and to complex geometries as well as the inclusion of a wide range of carrier dynamics as trapping and detrapping mechanisms. Moreover, since the method can provide the time evolution of the charge signal, it can be used to optimize the transfer function of the charge sensitive pre-amplifier.

**Acknowledgements**

The work of P. Olivero is supported by the "Accademia Nazionale dei Lincei – Compagnia di San Paolo" Nanotechnology grant, which is gratefully acknowledged.

**Figures**

**Fig. 1**

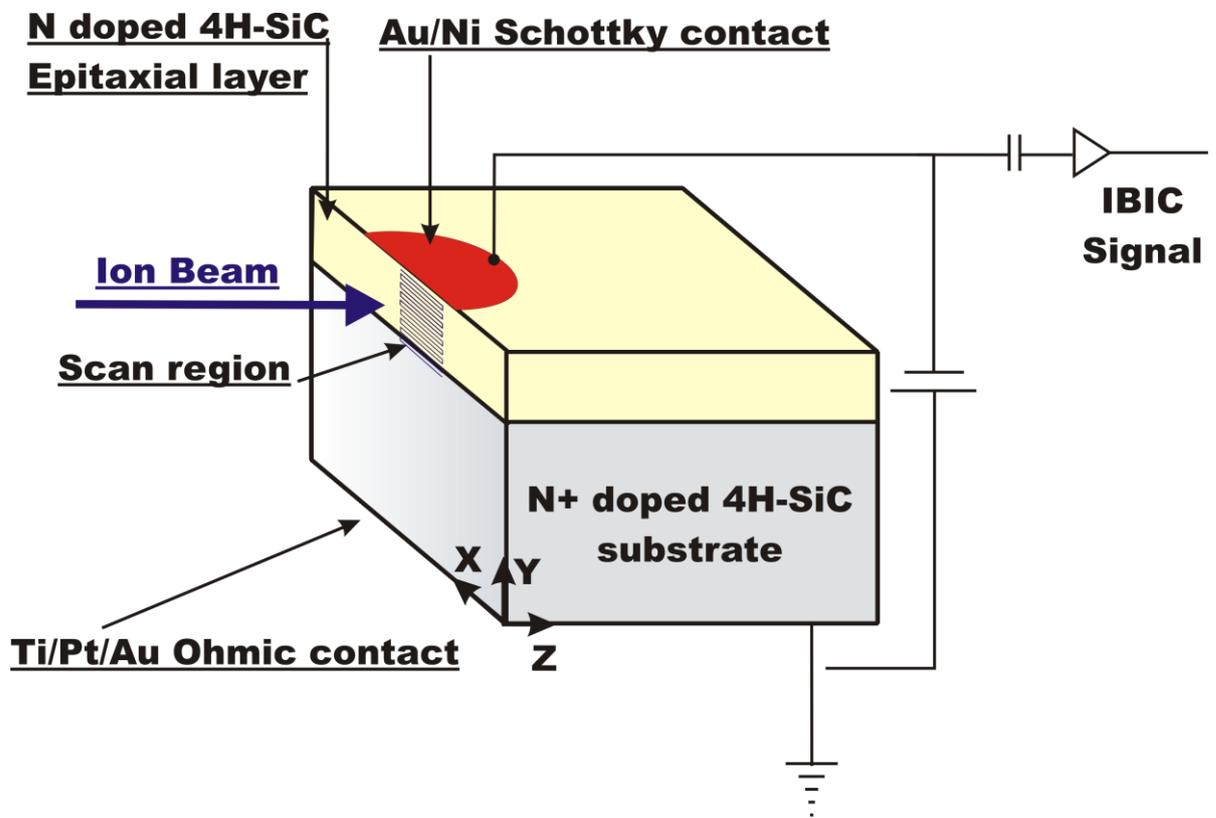

**Fig. 2**

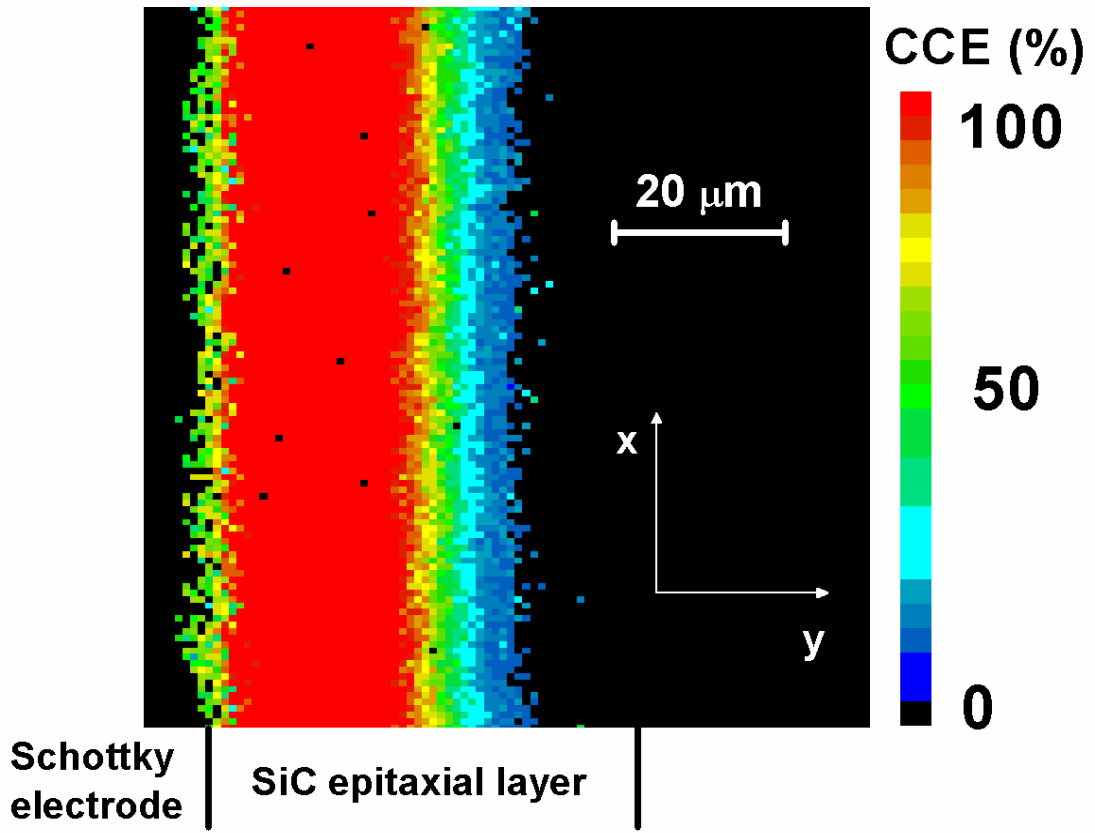

**Fig. 3**

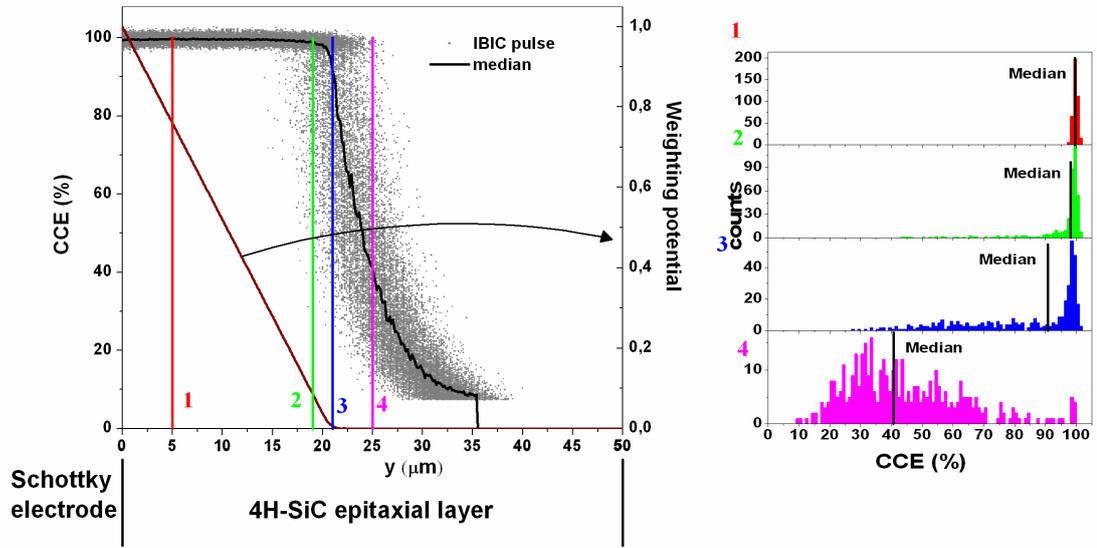

**Fig. 4**

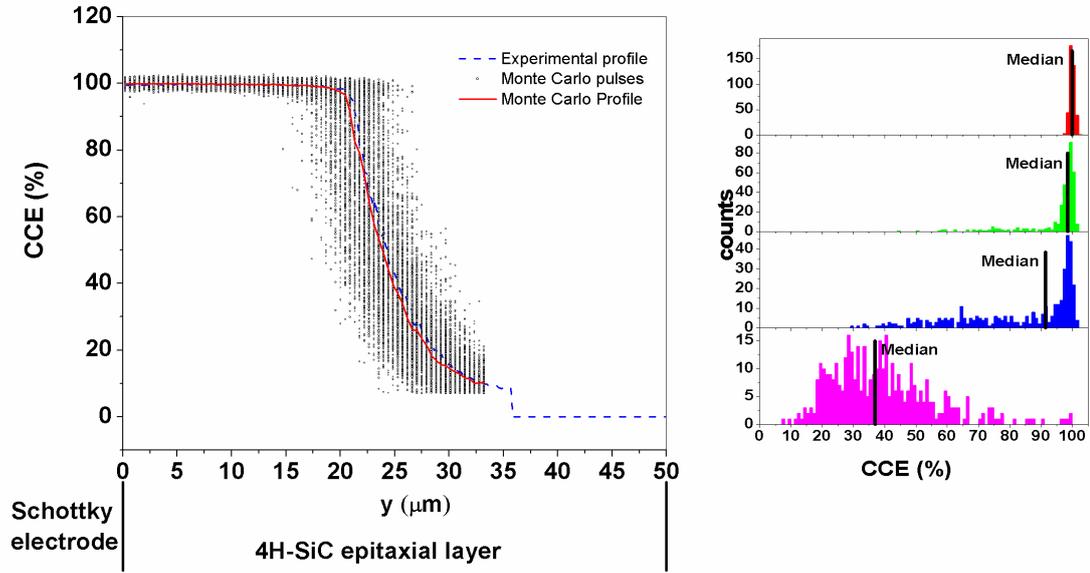

**Fig. 5**

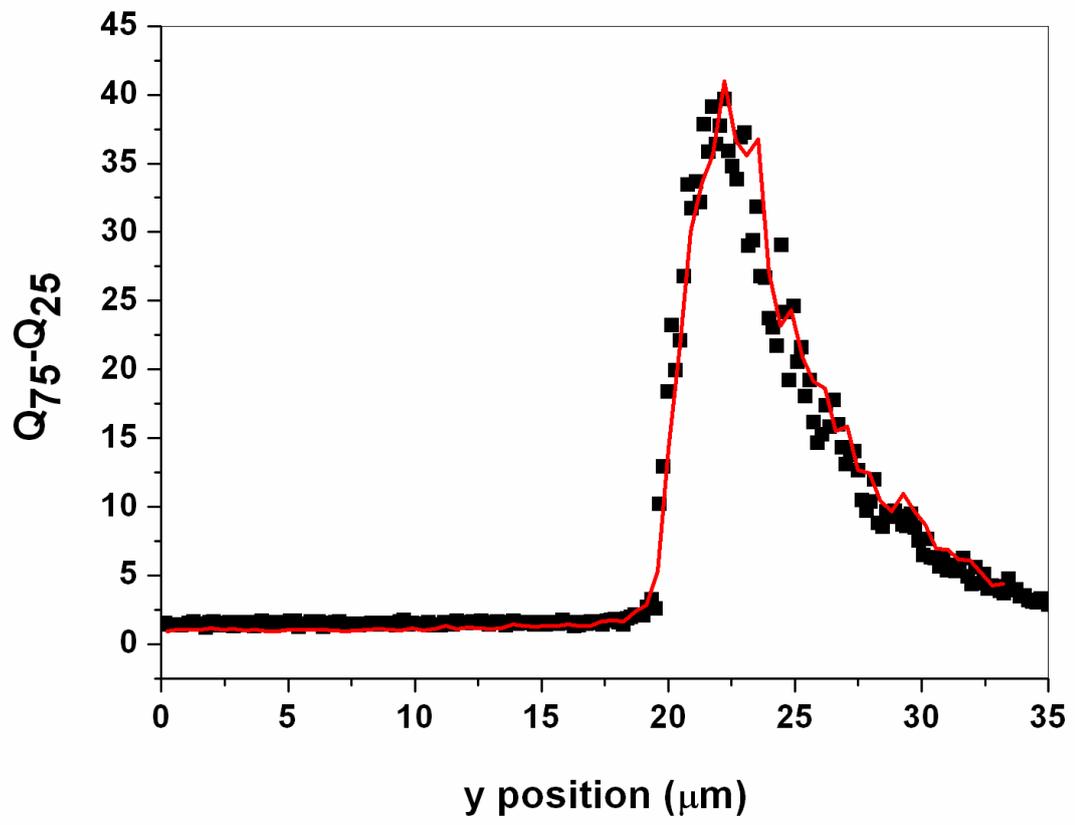

**Figure captions**

Fig. 1 (color online): Scheme of the lateral IBIC set-up.

Fig. 2 (color online): Lateral IBIC map at a reverse bias of 50 V.

Fig. 3 (color online): Left side: charge collection efficiency profile of the epitaxial layer of the 4H-SiC Schottky diode at a reverse bias voltage of 50 V. The dark red line represents the weighting potential (right scale). Right side: CCE spectra collected at the position indicated by the 4 lines in the figure on the left side.

Fig. 4 (color online): Simulated (continuous) and experimental (dashed) CCE median profiles at a reverse bias voltage of 50 V. Markers indicate the CCE for each simulated ion. On the right side, simulated CCE spectra collected at the position indicated by the 4 lines in the figure on the left side, to be compared with the experimental ones in Fig. 3.

Fig. 5 (color online): Comparison between the CCE spreads evaluated from Monte Carlo simulation (continuous line) and the experimental data (markers). The spreads are defined as the difference between the values of the third (Q75) and the first (Q25) quartile of the relevant CCE distributions.